\documentstyle[twoside,12pt]{article}
\newcommand{\be}{\begin{equation}}
\newcommand{\ee}{\end{equation}}
\newcommand{\bea}{\begin{eqnarray}}
\newcommand{\eea}{\end{eqnarray}}
\newcommand{\bdm}{\begin{displaymath}}
\newcommand{\edm}{\end{displaymath}}

\newcommand{\cod}{d^{\dagger}}

\newcommand{\we}{\wedge}

\newcommand{\om}{\omega}
\newcommand{\kh}{\hat{k}}

\begin{document}
\begin{titlepage}
\begin{center}
\vspace{1.8cm}
\vfill
{\large\bf A MASS BOUND FOR SPHERICALLY SYMMETRIC}
{\large\bf BLACK HOLE SPACETIMES}

\vspace{1.5cm}

{\bf{Markus Heusler}} \\
Enrico Fermi Institute, University of Chicago \\
5640 S. Ellis Ave., Chicago IL 60637 \\
\vspace{.3cm}
Nov. 15 1994

\end{center}
\vspace{1.5cm}

\begin{center}
{\bf Abstract}
\end{center}

\begin{quote}

Requiring that the matter fields are subject
to the dominant energy condition,
we establish the lower bound
$(4\pi)^{-1} \kappa {\cal A}$ for
the total mass $M$ of a static, spherically symmetric
black hole spacetime.
(${\cal A}$ and $\kappa$ denote the area
and the surface gravity of the horizon, respectively.)
Together with the fact that
the Komar integral provides a simple
relation between $M - (4\pi)^{-1} \kappa A$
and the strong energy condition, this enables
us to prove that
the Schwarzschild metric
represents the only static, spherically symmetric black
hole solution of a selfgravitating matter
model satisfying the dominant, but
violating the strong energy condition
for the timelike Killing field $K$ at every point,
that is, $R(K,K) \leq 0$.
Applying this result to scalar fields,
we recover the fact that
the only black hole configuration
of the spherically symmetric Einstein-Higgs model
with arbitrary non-negative potential is the
Schwarzschild spacetime with constant Higgs field.
In the presence of electromagnetic fields, we
also derive a stronger bound for the total mass,
involving the electromagnetic
potentials and charges.
Again, this estimate provides a simple tool to
prove a ``no-hair'' theorem for matter fields
violating the strong energy condition.

\end{quote}

\vfill

\end{titlepage}
%
\section{Introduction}

In 1979 Schoen and Yau succeeded
in proving the positive mass theorem by the means of a
variational technique \cite{SY}.
Subsequently, Witten presented a different proof
of the long standing problem, based on the
Lichnerowicz identity for spinor fields
\cite{Witt}, \cite{PT}.
Later on, Gibbons et al. have generalized Witten's
approach to black hole spacetimes, establishing
$M \geq 0$ for a spacelike hypersurface which is
regular outside an apparent horizon \cite{GHHP}.

According to Arnowitt, Deser and Misner (ADM), the
total energy (four-momentum) of an asymptotically
flat manifold can be defined in terms of a surface integral
at spacelike infinity, involving only the asymptotic behavior
of the metric \cite{ADM}.
Provided that spacetime admits a stationary symmetry,
Komar has given another expression for the total
mass in terms of the asymptotically timelike Killing field
\cite{Komar}.
With the help of Ricci's identity it is not difficult
to see that the positivity of the Komar mass is a
direct consequence of the {\em{strong}} energy condition
(SEC). Obviously this provides no alternative
proof of the positive
mass theorem for stationary spacetimes, since the SEC
is more restrictive than the {\em{dominant}} energy condition
(DEC) on which the positivity of the ADM mass is based.
However, if the matter model under consideration
satisfies the DEC but violates the SEC for
the timelike Killing field at every point of the domain,
$R(K,K) \leq 0$,
then the Komar expression
for the mass $M$ is non-positive, contradicting
the positive mass theorem, unless $M = 0$.
Hence, matter models of this kind admit only
trivial selfgravitating {\em{soliton}} solutions.

The question arises, whether this reasoning
can be extended to black hole spacetimes.
In this case, the
Komar integral yields the upper bound
$M \leq \frac{1}{4 \pi} \kappa {\cal A}$
if $R(K,K) \leq 0$, whereas the generalization
of the positive mass theorem still gives $M \geq 0$
as a consequence of the DEC \cite{GHHP}.
Hence, in order to establish "no-hair" results by applying
the above argument, it is desirable
to improve this bound and to conjecture that
for static black holes the DEC implies
$M \geq \frac{1}{4 \pi} \kappa {\cal A}$.

Until now, we did not succeed in deriving this
bound in the general static case.
However, restricting our attention to spherically symmetric
configurations, we present a simple proof of the
above inequality in this paper.

As an application, we show that
the only spherically symmetric
black hole solution of a static
selfgravitating harmonic map
with non-negative potential
and Riemannian target manifold
is the Schwarzschild metric
\cite{NMSA}, \cite{MSA}.
This generalizes the well-known result due to
Bekenstein \cite{Beke} to non-convex potentials.
As is clear from the above comments, the corresponding
conclusion can be drawn for {\em{soliton}} solutions
without any additional symmetry requirements.

In the presence of electromagnetic fields, stronger
estimates are needed in order to
take over the above reasoning, since the electromagnetic
part $T^{(em)}$ of the energy momentum tensor $T$
does not violate the SEC.
Taking advantage of Maxwell's equations and expressing
the total electric and magnetic charge
($Q$ and $P$)
in terms of volume integrals, it is not
difficult to show that
$M \leq \frac{1}{4 \pi} \kappa {\cal A} -
Q \Delta \phi - P \Delta \psi$, provided that
${\cal T}:= T - T^{(em)}$
violates the SEC for $K$ at every point of the domain
(where $\phi$ and $\psi$
denote the electric and the
magnetic potential, respectively).
Hence, in order to gain ``no-hair''
results in this case, one has to establish
the converse inequality on the basis of
the DEC.
We shall conclude this paper by
doing so for the spherically symmetric case.

\section{Komar Mass and Strong Energy Condition}

Throughout this article, we shall restrict ourselves to
strictly stationary, asymptotically flat spacetimes containing a
non-rotating black hole. Hence, we
assume that there
exists a Killing field $K$, timelike in all of the
domain of outer communications and
coinciding with the null-generator of the Killing
horizon $H$ (see \cite{HE} or \cite{CC}) for
details.
Taking advantage of the Komar integral
and Ricci's identity,
the goal of this section is
to establish the following
expression for the total mass in terms of the
$KK$-component of the Ricci tensor $R$,
\be
M \, = \, \frac{1}{4 \pi} \, \kappa \, {\cal A}
\, + \, \frac{1}{4 \pi} \, \int_{\Sigma}
\frac{R(K,K)}{V} \, \eta_{\Sigma} \, - \,
\frac{1}{2 \pi} \, \int_{\Sigma}
\frac{(\om,\om)}{V^2} \, \eta_{\Sigma} \, ,
\label{M1}
\ee
where $V$ and $\om$ denote the norm of the Killing field
and its twist $1$-form, respectively,
\be
V \, := \, - (K | K) \, \geq \, 0 \, ,
\; \; \; \; \;
\om \, := \, \frac{1}{2} \ast (K \we dK) \, .
\label{VW}
\ee
(Here and in the following
we use the symbol $K$ for both, the Killing
filed and its assigned $1$-form.)
$\eta_{\Sigma} := \ast K = i_K \eta$ denotes
the induced volume form on the spacelike
hypersurface $\Sigma$ extending from
spacelike infinity to the inner $2$-dimensional
boundary ${\cal H} = H \cap \Sigma$
(i.e., to the ``horizon at time $\Sigma$'').

As is seen from the above mass formula,
Einstein's equations
together with the SEC,
\be
T(\kh,\kh) \, + \, \frac{1}{2} \, tr(T) \, \geq \, 0 \, ,
\label{sec}
\ee
for the unit timelike field $\kh := K/\sqrt{V}$,
implies the positivity
of the quantity $M - (4 \pi)^{-1} \kappa {\cal A}$, provided
that spacetime is static, i.e., that the twist-form vanishes
identically. However, if the staticity requirement is dropped,
we are still able to conclude that the violation of the SEC
for the Killing field
$\kh$ at every point of the domain
implies the converse inequality,
$M \leq (4 \pi)^{-1} \kappa {\cal A}$.
(Note that in a strictly stationary domain $\om$ is
nowhere timelike and hence $(\om | \om) \geq 0$.)

In order to derive eq. (\ref{M1}) we recall that according
to Komar \cite{Komar} the mass of a stationary, asymptotically
flat spacetime can be expressed in terms of the asymptotically
timelike Killing field by the surface integral
\be
M \, = \,
- \frac{1}{8 \pi} \int_{S^2_{\infty}} \ast dK \, = \,
- \frac{1}{8 \pi} \int_{\cal H} \ast dK \, - \,
\frac{1}{8 \pi} \int_{\Sigma} d \ast dK \, ,
\label{M2}
\ee
where we have used Stokes' theorem in the second equality.
In the non-rotating case, to which we restrict our attention,
$K$ coincides with the null-generator Killing field of
the horizon \cite{HE}.

Denoting the second future directed null vector orthogonal
to the horizon with $n$ (normalized such that $(K|n) = -1$),
the boundary integral over ${\cal H}$
can be evaluated in the well known manner,
\be
\int_{\cal H} \ast dK \, = \,
\int_{\cal H} (K \we n | dK) \, d {\cal A} \, = \,
\int_{\cal H} (n | dV) \, d {\cal A} \, = \,
- 2 \kappa {\cal A} \, ,
\label{Ofl}
\ee
where we have used the definition $dV = 2 \kappa K$ of
the surface gravity $\kappa$ and the fact that the latter
is constant over the horizon (cf. eg. \cite{WK}).

In order to evaluate the volume integral in eq. (\ref{M2})
we take advantage of the Ricci identity for Killing fields,
\be
d \ast d K \, = \, 2 \ast R(K) \, ,
\label{Ric}
\ee
where the components of the Ricci $1$-form $R(K)$
are $R(K)_{\mu} = R_{\mu \nu} K^{\nu}$.
This yields the well-known formula
for the Komar Mass in terms of the horizon quantities
and the Ricci $1$-form,
\be
M \, = \, \frac{1}{4 \pi} \, \kappa {\cal A} \, - \,
\frac{1}{4 \pi} \int_{\Sigma} \ast R(K) \, .
\label{M3}
\ee

In order to proceed, we first recall the identity
\cite{NMFL}
\be
\ast d \om \, = \, - \, K \we R(K) \, ,
\label{dom}
\ee
which is obtained from the definition (\ref{VW})
of $\om$, the Ricci identity
(\ref{Ric}) and the fact that $\cod (K \we \Omega) =
-K \we \cod \Omega$ for an arbitrary $p$-form
$\Omega$ with vanishing Lie derivative with respect
to the Killing field K.
Making also use of the codifferential,
$\cod = \ast d \ast$,
and $\ast^2 \Omega = (-1)^{p+1} \Omega$, we have
$\ast d \om = (1/2) \cod (K \we dK) =
-(1/2) K \we \cod d K = - K \we R(K)$, completing
the derivation of eq. (\ref{dom}).
Applying the inner derivation $i_K$ on the
above identity and recalling that
$i_K \ast \Omega \equiv (-1)^p \ast (K \we \Omega)$,
we immediately obtain
\be
K \we d \om \, = \, V \ast R(K) \, + \,
R(K,K) \ast K \, ,
\label{ID1}
\ee
which enables us to express the integrand
in eq. (\ref{M3}) in terms of $R(K,K)$:
\be
M \, = \, \frac{1}{4 \pi} \, \kappa {\cal A} \, + \,
\frac{1}{4 \pi} \int_{\Sigma} \frac{R(K,K)}{V} \ast K \, - \,
\frac{1}{4 \pi} \int_{\Sigma} \frac{K \we d \om}{V} \, .
\label{M4}
\ee

This already proves eq. (\ref{M1}) in the static case.
However, if $\om \neq 0$, the last term
in the above equation needs further transformations.
Noting that the definition (\ref{VW}) of the twist-form
implies $d (K/V) = - (2/V^2) i_K \ast \om$,
it is easy to derive the useful identity
\be
d (\frac{K \we \om}{V}) \, = \,
\frac{2}{V^2} (\om | \om) \ast K
\, - \, \frac{1}{V} K \we d \om \, .
\label{ID2}
\ee
Integrating this identity over $\Sigma$
and making use of Stokes'
theorem and the fact that the integrand $(K \we \om)/V$
vanishes on both components of the boundary, $S^2_{\infty}$
and ${\cal H}$, we have
\be
\int_{\Sigma} \frac{K \we d \om}{V} \, = \,
2 \int_{\Sigma} \frac{(\om | \om)}{V^2} \ast K \, ,
\label{ID3}
\ee
which completes the proof of the mass formula (\ref{M1}).

It is also worth noticing that the identity
(\ref{ID2}) yields a very direct proof of the fact that a
strictly stationary spacetime is static, if it is Ricci static:
First of all, eq. (\ref{dom}) shows that
$R(K) \we K = 0$ (Ricci-staticity) is equivalent to
$d\om = 0$. The integrated version (\ref{ID3}) of
the identity (\ref{ID2}) then implies that $\om$
vanishes itself. The only non-trivial task
is to establish that the boundary
term does not contribute
on the horizon \cite{NMFL}, which can be concluded
from the general properties of Killing
horizons \cite{CC}.
(We point out that the validity of this simple
staticity proof is restricted to the case where
the Killing field is timelike in all
of the domain (strict stationarity).
In order to overcome this difficulty,
one takes advantage of a recent theorem by
Chru\'sciel and Wald \cite{PCRW}, establishing
the existence of a maximal slice.)

To summarize, we note
that in a strictly stationary,
asymptotically flat black hole spacetime the violation
of the SEC for the unit Killing field at every point
$\kh = K/\sqrt{V}$ implies an
upper bound for the total mass in terms of the
horizon quantities $\kappa$ and ${\cal A}$, that is,
\be
M \, \leq \; \frac{1}{4 \pi} \kappa {\cal A} \, , \;
\; \; \; \mbox{if} \; \; \;
T(\kh,\kh) \, + \, \frac{1}{2} \, tr(T) \, \leq \, 0 \, ,
\label{upper}
\ee
whereas the converse inequality holds
if the SEC is fulfilled and the domain is static.

\section{Dominant Energy Condition and Spherical Symmetry}

The violation of the SEC for the timelike Killing field
implies that the total mass cannot exceed
the value $\frac{1}{4 \pi} \kappa {\cal A}$
of the Komar integral at the horizon.
Considering static spacetimes containing no black hole region,
we immediately obtain $M \leq 0$, contradicting
the positive energy theorem which is based on the DEC.
This enables us to conclude that there are no non-trivial
self-gravitating {\em{soliton}} solutions for matter models
obeying the DEC but violating the SEC for the Killing
field $\kh$ at every point.

The extension of the argument to spacetimes
containing a black hole would require a proof
of the inequality $M \geq \frac{1}{(4 \pi)} \kappa {\cal A}$
as a consequence of the DEC.
However, to our knowledge, a conjecture of this
kind has not been established until now.
In the general case, the strongest result
consists in the generalization of Witten's
proof of the positive
mass theorem \cite{Witt} (cf. also \cite{PT})
by Gibbons et. al. \cite{GHHP},
establishing $M \geq 0$ for
black hole spacetimes as well.

As we shall demonstrate in this section it is,
however, not hard to derive the desired bound
in the spherical symmetric case.
As a consequence, we conclude that all
static, spherically symmetric black hole
solutions with matter satisfying the DEC but
violating the SEC for $\kh$ coincide with
the Schwarzschild metric.

A static and spherically symmetric metric is
parametrized by two functions depending on the
radial coordinate $r$ only. Using the quantities
$m(r)$ and $\delta (r)$ introduced in \cite{NMSA}
we write
\be
g \, = \, - N S^2 \, dt^2 \, + \, N^{-1} \, dr^2
\, + \, r^2 \, d \Omega ^2 \, ,
\label{metric}
\ee
\be
N(r) \, = \, 1 - \frac{2 m(r)}{r} \, ,
\; \; \; \; \; \;
S(r) \, = \, e^{- \delta(r)} \, .
\label{ndelta}
\ee
(In terms of the familiar Schwarzschild parametrization
one has $\delta = -(a+b)$ and $N = e^{-2b}$.)
Tensor components will refer to the orthonormal
frame $\{ \theta^{\mu} \}$ of $1$-forms,
\be
\theta^0 \, = \, \sqrt{N} S \, dt \, ,
\; \; \; \;
\theta^1 \, = \, \frac{1}{\sqrt{N}} \, dr \, ,
\; \; \; \;
\theta^2 \, = \, r \, d \vartheta \, ,
\; \; \; \;
\theta^3 \, = \, r \, \sin \vartheta \, d \varphi \, ,
\label{frame}
\ee
where $S$ is positive and so is $N$, except on the
horizon, $r = r_H$, where $N$ vanishes by
definition.
In terms of these quantities the Killing $1$-form
becomes $K = -N S^2 dt = - \sqrt{N} S \theta^0$
and thus, since $\ast (\theta^0 \we \theta^1) =
- \theta^2 \we \theta^3$, we have
$\ast dK = S^{-1} \frac{d(NS^2)}{dr}
\ast (\theta^0 \we \theta^1) =
-r^2 S^{-1} (NS^2)' d \Omega$.
Defining the ``local mass'' $M(r)$ by the
Komar integral over a $2$-sphere with
coordinate radius $r$ now yields
\be
M(r) \, = \, - \frac{1}{8 \pi} \int_{S^2_r} \ast dK \, = \,
\frac{r^2}{2 S} (NS^2)' \, = \, mS \, + \, N S' r^2 \, - \,
m' S r \, .
\label{local}
\ee
As we have argued in the previous section,
$M(r)$ is a decreasing function if $R(\kh , \kh)$
is negative, i.e., if the SEC is violated for the
Killing field $\kh$.
However, as a consequence of the DEC, we shall
now establish that
$\lim_{\infty} M(r) \geq M_H$,
independently
of whether or not the SEC holds.

Computing the components of the Einstein tensor
in the orthonormal frame (\ref{frame}), one finds
\be
G_{00} \, = \, \frac{2}{r^2} \, m' \, ,
\; \; \; \; \; \;
G_{00} \, + \, G_{11} \, = \, \frac{2 N}{r} S^{-1} \, S' \, .
\label{Einstein}
\ee
Requiring that $T(X)$ is future directed timelike or null
for all future directed timelike vectors $X$, Einstein's
equations imply that the quantities
$G_{00}$ and $G_{00} + G_{11}$
are not negative.
Hence, provided that matter is subject to the DEC,
both $m(r)$ and $S(r)$ are increasing functions,
\be
m'(r) \, \geq \, 0 \, ,
\; \; \; \; \; \;
S'(r) \, \geq \, 0  \,
\; \; \; \mbox{for} \; \; \;
r \, \geq \, r_H \, .
\label{increase}
\ee

In order to gain estimates
for $M(r)$ at infinity and at the
horizon, it remains to note that the last
term in eq. (\ref{local}) does not contribute
as $r \rightarrow \infty$, whereas
the second term vanishes at the horizon.
We obtain thus an upper bound for $M_H$ and
a lower bound for $M := \lim_{\infty} M(r)$.
More precisely, asymptotic flatness
implies the existence and finiteness of
$m_{\infty} := \lim_{\infty} m(r)$,
$S_{\infty} := \lim_{\infty} S(r)$ and
$\lim_{\infty} (r^2 S')$.
Since, in addition,
$\lim_{\infty} (r m')$ vanishes and the
second term in eq. (\ref{local}) is non-negative, we have
$M \geq m_{\infty} S_{\infty}$.
For $r = r_H$, $N(r)$ vanishes which,
together with the fact that $m' S r$ is
non-negative yields the estimate
$M_H \leq m_H S_H$ for the Komar
integral at the horizon.
Hence, taking again advantage of the circumstance
that $m(r)$ and $S(r)$ are increasing,
we find
\be
M_H \, \leq \, m_H S_H \, \leq \,
m_{\infty} S_{\infty} \, \leq \, M \, ,
\label{ineqs}
\ee
which establishes the fact that the difference of the
Komar integrals is not negative if the
matter model is subject to the DEC,
\be
- \frac{1}{8 \pi} \int_{\partial \Sigma}
\, \ast dK \, = \,
M \, - \, M_H \, = \,
M \, - \, \frac{1}{4 \pi} \kappa {\cal A} \, \geq \, 0 \, ,
\label{final}
\ee
where $\partial \Sigma = S^2_{\infty} - {\cal H}$.

As a consequence, we conclude that both functions,
$m(r)$ and $S(r)$, have to assume constant values if the
DEC holds and the
SEC is violated for the timelike Killing field.
In this case, the metric (\ref{metric})
coincides with the Schwarzschild metric and the energy
momentum tensor thus vanishes.

It is probably worth noting that the DEC and eq. (\ref{local}) together
provide a simple estimate for the surface gravity of a
spherically symmetric black hole:
Evaluating the formula for $M(r)$ at the horizon, and using
the fact that $M_H = \frac{1}{4 \pi} \kappa {\cal A} = \kappa r^2_H$,
immediately yields
\be
\kappa \, = \, \frac{S_H}{2 r_H} \, (1 \, - \, 2 m'_H) \, .
\label{star}
\ee
Since the DEC implies that $m'(r) \geq 0$
and $S_H \leq S_{\infty} = 1$, we recover the fact
\cite{Visser}, that the
Hawking temperature $T_H = \kappa \hbar /(2 \pi k)$
of a non-degenerate, spherically
symmetric black hole with matter satisfying the DEC is
bounded from above by the Hawking temperature $T_H^{(vac)}$
of a Schwarzschild black hole (with the same area),
\be
T_H \, = \, \frac{\hbar}{2 \pi k}
\, \frac{S_H}{2 r_H} \, (1 \, - \, 2 m'_H) \, \leq \,
\frac{\hbar}{2 \pi k}
\, \frac{1}{2 r_H} \, = \, T_H^{(vac)} \, .
\label{TH}
\ee

\section{Application to Higgs Fields and Harmonic Maps}

As already mentioned,
the above result provides a very
simple proof of the "no-hair" theorem for black
hole solutions of self-gravitating
spherically symmetric Higgs fields with
arbitrary non-negative potentials.
This result was
already derived by Straumann and the author
by means of scaling
arguments \cite{NMSA}, \cite{MSA}.
A different proof, taking advantage
of the existence of a monotonic function
was recently given by Sudarsky \cite{Suda}.
As a matter of fact, it is exactly the violation
of the SEC which renders possible the construction
of a Liapunov function, since
eq. (\ref{Ric}) becomes for scalar fields
with potential $P[\phi]$
\be
d \ast dK \, = \, 16 \pi \, P[\phi] \, \ast K
\; \; \; \Rightarrow \; \; \;
[r^2 S^{-1} \, (N S^2)']' \, = \, 16 \pi \, S \, P[\phi] \,
\geq \, 0 \, .
\label{sud}
\ee

In order to apply the results obtained above,
it only remains to verify that in
a static domain, a
selfgravitating harmonic map
(non-linear $\sigma$-model)
with an additional non-negative potential term satisfies the
DEC and violates the SEC for the timelike Killing field
at every point.
We recall that a mapping
$\phi$ from spacetime $(M,g)$ into a
Riemannian manifold $(N,G)$ is said to be
harmonic, if it is a solution to the variational
equations for the matter Lagrangian
\be
\frac{1}{2} G_{AB}(d \phi^A | d \phi^B) \, = \,
\frac{1}{2} G_{AB}[\phi(x)] \, g^{\mu \nu}(x)
\partial_{\mu} \phi^A \, \partial_{\nu} \phi^B) \, ,
\label{harm}
\ee
where $G[\phi]$ denotes the Riemannian metric of the target
manifold $N$ \cite{EL}. In the simplest case, that is, if
the target manifold is assumed to be a vector space,
this reduces to the ordinary scalar field
Lagrangian, describing a set of Higgs fields,
provided that an additional potential is taken into account
as well.

Hence, we consider the matter Lagrangian
\be
{\cal L} \, = \,
\frac{1}{2} G_{AB}(d \phi^A | d \phi^B) \, + \,
P[\phi] \, ,
\label{lag}
\ee
with arbitrary Riemannian
target metric $G[\phi]$ and arbitrary
non-negative potential $P[\phi]$.
The energy momentum tensor becomes
\be
T \, = \, G_{AB} \, d \phi^A \otimes d \phi^B \, - \, g \, {\cal L} \, .
\label{enmo}
\ee
Together with the requirement that $\phi$ is static,
i.e., $L_K \phi^A = 0$, we immediately have
$T(\kh , \kh) = {\cal L}$ and $tr(T) = -2 {\cal L} - 2 P[\phi]$,
implying the violation of the SEC for $\kh$ at every
point of the domain,
\be
T(\kh,\kh) \, + \, \frac{1}{2} \, tr(T) \, =
\, - P[\phi] \, \leq \, 0 \, .
\label{sec1}
\ee
On the other hand, the energy momentum tensor (\ref{enmo})
clearly fulfills the DEC, since
\be
T_{00} \, = \, {\cal L} \, ,
\; \; \; \; \; \;
T_{00} + T_{11} \, = \,
G_{AB} (d \phi^A)_1 (d \phi^B)_1 \, = \,
G_{AB} \, N \, (\phi^A)' \, (\phi^B)' \, .
\label{decsc}
\ee
Since the violation of the SEC for $\kh$
implies that $M \leq \frac{1}{4 \pi} \kappa {\cal A}$,
whereas the DEC implies the converse inequality, we
obtain $M = \frac{1}{4 \pi} \kappa {\cal A}$ and
thus $m' = S' = 0$ and $\phi' = 0$, $P[\phi] = 0$.
Hence, the metric is the Schwarzschild metric and
the scalar fields have to assume a vacuum configuration.

\section{Electromagnetic Fields}

Let us now consider the case where a part of
the matter consists of electromagnetic
fields.
Let $T_{\mu \nu} = T^{(em)}_{\mu \nu} +
{\cal T}_{\mu \nu}$, where
\be
T^{(em)}_{\mu \nu} \, = \, \frac{1}{4 \pi} \,
[F_{\mu}^{\; \sigma} F_{\nu \sigma} - \frac{1}{2}
g_{\mu \nu}  (F | F)]
\label{EM1}
\ee
and ${\cal T}_{\mu \nu}$ denotes the energy-momentum
tensor of the remaining matter fields. It is easy to
see that
\be
8 \pi \, T^{(em)}(K,K) = (E|E) \, + \, (B|B) \, ,
\label{EM2}
\ee
where the $1$-forms $E$ and $B$ are defined with
respect to the Killing field,
$E := -i_K F$, $B := i_K \ast F$, also implying
the identity $V (F|F) = (B|B) - (E|E)$.
Restricting ourselves to static configurations
(the generalization to $\omega \neq 0$ is straightforward),
eq. (\ref{M1}) becomes
\be
M \, = \, \frac{1}{4 \pi} \, \kappa \, {\cal A}
\, + \, \frac{1}{4 \pi} \, \int_{\Sigma}
\frac{(E|E) + (B|B)}{V} \, \eta_{\Sigma} \, + \,
2 \, \int_{\Sigma}
[ {\cal T}( \kh , \kh ) + \frac{1}{2} tr ({\cal T})]
\, \eta_{\Sigma} \, ,
\label{EM3}
\ee
which shows that
${\cal T}( \kh , \kh ) + \frac{1}{2} tr ({\cal T}) \leq 0$
does no longer imply
$M \leq \frac{1}{4 \pi} \kappa {\cal A}$.
However, as is well known, the electromagnetic
contributions can be transformed into surface terms.
In order to do this, we recall that the static Maxwell
equations assume the simple form (cf. eg. \cite{MH94})
\be
dE \, = \, dB \, = \, 0 \, ,
\; \; \; \; \;
\cod (E/V) \, = \, \cod (B/V) \, = \, 0 \, ,
\label{EM4}
\ee
which also implies the existence of the two
potentials $\phi$ and $\psi$, with $E = d \phi$ and $B = d \psi$.
(Here we assume that
$\Sigma$ is simply connected, i.e., that
each component of the
horizon at time $\Sigma$ is a topological $2$-sphere
(cf. \cite{HE}, and \cite{NEW} for new results).)

Integrating $d \ast F = 0$ over $\Sigma$, one
immediately finds that the surface integrals of
$\ast F$ over the
horizon and over $S^2_{\infty}$ are equal, the latter being
$-4 \pi$ times the electric charge $Q$ by definition.
Taking also advantage of
the fact that in the static case $E \we B =0$,
we have $d (\phi \ast F) = E \we \ast F =
V^{-1}[E \we \ast(K \we E)] = V^{-1}(E|E) \ast K$,
which yields
\be
-4 \pi Q \Delta \phi \, = \,
\int_{\partial \Sigma} \phi \ast F \, = \,
\int_{\Sigma} \frac{(E|E)}{V} \ast K \, ,
\label{EM6}
\ee
where $\Delta \phi := \phi_{\infty} - \phi_{H}$ and where
we have also used the fact that the electric potential
assumes a constant value over the horizon
(cf. eg. \cite{NMFL}).
Since similar considerations apply to the magnetic field,
one obtains the generalized Smarr formula
\be
M \, - \, \frac{1}{4 \pi} \kappa {\cal A}
\, + \, Q \Delta \phi \, + \,
P \Delta \psi \, = \,
2 \, \int_{\Sigma}
[ {\cal T}( \kh , \kh ) + \frac{1}{2} tr ({\cal T})]
\, \eta_{\Sigma} \, ,
\label{Smarr}
\ee
Hence, the combined Komar integral $\tilde{I}$,
\be
\tilde{I} \, := \, - \frac{1}{4 \pi}
\int_{\partial \Sigma} \ast d \tilde{K} \, ,
\; \; \;
\mbox{where} \; \; \;
d \tilde{K} \, = \,
dK \, + \, 2 \phi F \, - \, 2 \psi \ast F
\label{Comb}
\ee
is non-positive whenever the additional matter fileds
violate the SEC for the timelike Killing field
at every point of the domain.
(Note that $2 \phi F - 2 \psi \ast F$ is closed, which
justifies the introduction of the $1$-form $\tilde{K}$.)
It is also instructive to compute the differential
of $\ast d \tilde{K}$,
\bdm
d \ast d \tilde{K} \, = \, d \ast dK \, + \,
2 \, (E \we \ast F + B \we F)
\edm
\bdm
= \, 2 \ast \, [R(K) \, - \, 8 \pi T^{(em)}(K)] \, = \,
16 \pi \, \ast
[ {\cal T}(K) - \frac{1}{2} tr({\cal T}) K ] \, ,
\edm
which provides a second derivation of the mass
formula (\ref{Smarr}), since for static configurations
${\cal T}(K,K) \ast K = -V \ast {\cal T}(K)$.
Hence, in the presence of electromagnetic fields,
the estimate (\ref{upper}) given at the end
of the second section generalizes to
\be
M \, \leq \, \frac{1}{4 \pi} \kappa {\cal A}
\, - \, Q \Delta \phi \, - \, P \Delta \psi \, , \;
\; \; \; \mbox{if} \; \; \;
{\cal T}(\kh,\kh) \, + \, \frac{1}{2} \, tr({\cal T}) \, \leq \, 0 \, .
\label{upperem}
\ee

The objective is now to establish the converse inequality
on the basis of the DEC.
As in the third section, we are able to reach this goal only
under the additional assumption of spherical symmetry.
With respect to the orthonormal tetrad introduced in
eq. (\ref{frame}) we have
\be
T^{(em)}_{00} \, = \, - \, T^{(em)}_{11} \, = \,
\frac{1}{8 \pi} \, S^{-2} \, (\phi'^2 \, + \, \psi'^2) \, .
\label{EM10}
\ee
Hence, Einstein's equations, together with the general
identities (\ref{Einstein})
and the dominant energy condition for ${\cal T}$
imply the inequalities
\be
2 r^{-2} \, m' \, - \,
S^{-2} \, (\phi'^2 \, + \, \psi'^2) \, \geq \, 0 \, ,
\; \; \; \; \;
N r^{-1} \, S^{-1} S' \, \geq \, 0 \, .
\label{INEQs}
\ee
In addition, in the spherically symmetric case,
Maxwell's equations for the potentials
$\phi$ and $\psi$ simply reduce to
\be
Q \, = \, - \, r^2 S^{-1} \phi ' \, ,
\; \; \; \; \;
P \, = \, - \, r^2 S^{-1} \psi ' \, .
\label{MaxSpher}
\ee
Our task is now to show that these relations imply the
converse estimate for $M$ than
the one given in eq. (\ref{upperem}),
that is, we have to establish that the combined Komar integral
$\tilde{I}$ is non-negative.
Using the expression (\ref{local}), we obtain
\be
\tilde{I} \, = \,
\Delta (mS) \, + \, \Delta (N S' r^2) \, - \,
\Delta (m' S r) \, + \, Q \Delta \phi \, + \,
P \Delta \psi \, ,
\label{Itil}
\ee
$\Delta$ denoting the difference of the quantities between
$S^2_{\infty}$ and the horizon.
Since $\lim_{\infty} (S r) = 0$, $S' \geq 0$ and
$N_H = 0$, we have
$\Delta (N S' r^2 - m' S r) \geq m'_H S_H r_H$,
which we use to write
\be
\tilde{I} \, \geq \,
S_{\infty} (\Delta m + m_H) \, - \,
S_H (m_H - m'_H r_H) \, + \,
Q \Delta \phi \, + \, P \Delta \psi \, .
\label{Itil1}
\ee
In order to proceed, we need estimates for
$Q \Delta \phi$, $P \Delta \psi$ and $\Delta m$.
First of all, integrating Maxwell's equations
(\ref{MaxSpher}) and using the fact
that $0 \leq S \leq S_{\infty}$ immediately yields
\be
Q \Delta \phi \, \geq \, - S_{\infty} \frac{Q^2}{r_H} \, ,
\; \; \; \; \;
P \Delta \psi \, \geq \, - S_{\infty} \frac{P^2}{r_H} \, .
\label{ineq1}
\ee
Secondly, inserting Maxwell's equations into the
inequality (\ref{INEQs}) for $m'$
and integrating again, we also have
\be
m' \, \geq \frac{Q^2 + P^2}{2 r^2} \, ,
\; \; \; \; \;
\Delta m \, \geq \frac{Q^2 + P^2}{2 r_H} \, .
\label{ineq2}
\ee
Taking advantage of these relations, we finally obtain
the following lower bound for the combined Komar integral
$\tilde{I}$
\bea
\tilde{I} & \geq &
S_{\infty} (\frac{Q^2 + P^2}{2 r_H} + m_H) +
S_H (\frac{Q^2 + P^2}{2 r_H} - m_H) -
S_{\infty} \frac{Q^2 + P^2}{r_H} \\
\nonumber
& = &
\Delta S (m_H - \frac{Q^2 + P^2}{4 m_H}) \, \geq \, 0 \, .
\label{Ibound}
\eea
Here we have used the regularity property of
the horizon to perform the last step:
As a matter of fact,
taking advantage of the estimate (\ref{INEQs}) for $m'$
as well as of the expression (\ref{local})
at $r = r_H$, we obtain
\bdm
m_H - \frac{Q^2 + P^2}{4 m_H} \, \geq \,
S_H^{-1} (m_H S_H - m'_H S_H r_H) \, = \,
S_H^{-1} M_H \, = \, \frac{1}{4 \pi S_H} \kappa {\cal A} \, \geq \, 0 \, .
\edm
This completes the demonstration that spherically symmetric
black hole configurations with electromagnetic fields and
matter satisfying the DEC are subject to the inequality
\be
M \, \geq \, \frac{1}{4 \pi} \kappa {\cal A}
\, - \, Q \Delta \phi \, - \, P \Delta \psi \, .
\label{lowerem}
\ee

Hence, if ${\cal T}$ violates the SEC at every point
for the timelike
Killing, the relations (\ref{upperem}) and
(\ref{lowerem}) imply that
the inequalities (\ref{INEQs})
become equalities. Computing the metric
functions $S$ and $m$ and using the definition
$N(r) = 1 - 2m(r)/r$ then yields the result
\be
S \, = \, const \, ,
\; \; \; \;
N \, = \, 1 - \frac{2m_{\infty}}{r} + \frac{Q^2 + P^2}{r^2} \, ,
\; \; \; \;
\phi \, = \, \frac{Q}{r} \, ,
\; \; \; \;
\psi \, = \, \frac{P}{r} \, ,
\label{RN}
\ee
i.e., the Reissner-Nordstr\"om solution, where
the remaining matter fields have to assume a vacuum configuration,
${\cal T}_{\mu \nu} = 0$.
As an application, this enables us to
generalize the ``no-hair'' result for selfgravitating
scalar fields discussed in the previous section
to the case where electromagnetic fields are present as well.

We finally note that the estimate (\ref{star})
for the Hawking temperature
together with the inequality (\ref{ineq2}) for $m'$ now yields
the stronger bound \cite{Visser}
\be
T_H \, = \, \frac{\hbar}{2 \pi k}
\, \frac{1}{2 r_H} \, (1 \, - \, \frac{Q^2 + P^2}{r_H^2})
\, = \, T_H^{(evac)} \, ,
\label{TH2}
\ee
$T_H^{(evac)}$ denoting the Hawking temperature of
the Reissner-Nordstr\"om black hole.

\section{Conclusion}

We have established the lower bounds
(\ref{final}) and (\ref{lowerem})
for the total mass of a spherically symmetric
black hole spacetime with matter satisfying the DEC.
On the other hand, for a static
(strictly stationary) configuration the
Komar expressions for the mass and
electromagnetic monopole charges imply
the converse estimates, provided that
$T - T^{(em)}$ violates the SEC for the
timelike Killing field $K$ at every point of the domain.
This yields a simple criterion
which, for instance, enables one to
exclude black hole solution with scalar hair.

As already mentioned, this "no-hair" theorem
has been established some
time ago by Bekenstein \cite{Beke} and generalized to arbitrary
non-negative potentials by different means.
However, in contrast to our earlier attempts,
\cite{NMSA}, \cite{MSA} the
reasoning presented here enables one to
exclude non-trivial solutions by simply verifying
the energy conditions.
Unfortunately, the second part of the argument, i.e.,
the proof that the mass can be estimated from below by
the Komar integral over the horizon, is heavely based
on the requirement of spherical symmetry.
We think that it might be interesting to
investigate whether this bound can be derived
on the basis of staticity alone,
that is, without assuming spherical symmetry.

If no potential is taken into account, we
were able to establish the general "no-hair"
theorem for selfgravitating static harmonic maps
\cite{Msigma} by adapting the uniqueness proof for static
vacuum black holes \cite{BM}.
It is also possible to handle the
case of a convex potential, by
taking advantage of Stoke's theorem \cite{Beke}
(see also \cite{GG}). However, without imposing
spherical symmetry,
the problem still remains open for
arbitrary non-negative potentials.

We finally point out that all uniqueness
results for scalar maps
rely on the {\em{harmonic}} structure of the
Lagrangian. This is a necessary restriction,
as is reflected by the fact that
there do
exist black hole solutions with scalar hair
if more general actions are admitted
\cite{DHS}, \cite{HSZ}.

I wish to thank Robert Wald for helpful
discussions and comments and Norbert
Straumann for reading the manuscript and
suggesting some improvements.
I also wish to thank the Enrico Fermi Institute
for its hospitality. This work was supported by the
Swiss National Science Foundation.

\end{document}